\documentclass[twocolumn,aps]{revtex4}

\usepackage{amsmath}
\usepackage{graphicx}

\def\){\right)}
\def\({\left(}
\def\]{\right]}
\def\[{\left[}

\newcommand{\be}{\begin{equation}}
\newcommand{\ee}{\end{equation}}

\newcommand{\roughly}[1]%
{\mathrel{\raise.4ex\hbox{$#1$\kern-.75em\lower1ex\hbox{$\sim$}}}}

\newcommand\beq{\begin{eqnarray}}\newcommand\eeq{\end{eqnarray}}

\def\Dsl{\,\raise.15ex \hbox{/}\mkern-12.8mu D}

\def\fm3{fm$^{-3}$}

\begin{document}

\title{Leading Order $k_Fa$ Corrections to the Free Energy and Phase Separation in Two-component Fermion Systems}

\author{Heron Caldas}

\affiliation{Departamento de Ci\^{e}ncias Naturais, Universidade Federal de S\~{a}o Jo\~{a}o Del Rei, Pra\c{c}a Dom Helv\'{e}cio 74, 36301-160, S\~{a}o Jo\~{a}o Del Rei, MG, Brazil \\ }

\begin{abstract}
We study phase separation in a dilute two-component Fermi system with attractive interactions as a function of the coupling strength and the polarization or number density asymmetry between the two components. In weak and strong couplings with a finite number density asymmetry, phase separation is energetically more favorable.  A heterogeneous phase containing a symmetric superfluid component and an asymmetric normal phase has lower energy than a homogeneous normal phase. We show that for a small number density asymmetry, taking into consideration the leading order corrections at order $k_Fa$ of the interaction parameter, phase separation is stable against the normal phase in the whole BCS range. We investigate the consequences of the consideration of the leading order $k_Fa$ corrections to the thermodynamic potentials of the normal and BCS phase on the Chandrasekhar-Clogston limit. We have also investigated the stability of a Bose-Fermi mixture in the far-BEC limit. We find that the molecular BEC is locally stable against an external magnetic field $h$, provided $|h|$ is smaller than the pairing gap $\Delta_{gap}$.

\end{abstract}
\pacs{74.20.Fg,03.75.Ss,21.65.+f,}
\maketitle

\section{Introduction}

It is well known that attractive interactions among fermions at sufficiently low temperature destabilize the Fermi surface. This instability, which is successfully explained by the Bardeen-Cooper-Schrieffer (BCS) theory of superconductivity, is characterized by pairing between spin-up and spin-down particles with opposite momenta near their common Fermi surface and results in the appearance of superfluid properties by the system. Besides, there is also the emergence of an energy gap in the excitation spectrum. Recent experiments on cold fermionic atoms, demonstrating an enormous ability to tune several physical parameters in a broad range, such as temperature, number density of different species (spin-up $\equiv \uparrow$ and spin-down $\equiv \downarrow$) and atom-atom interaction~\cite{Thomas:2004ex,Bartenstein:2004,Chin:2004,Greiner:2004,Review}, have motivated a great theoretical interest in fermion superfluids~\cite{Review,Theory1,Theory2,Zwerger}.

The pairing in spin-polarized systems, where there is a mismatch in the two-species Fermi surfaces, raises the possibility of unconventional and even exotic phases since this unfavorable situation precludes the system to have a standard BCS ground state. Several candidates have been proposed as, for example, a gapless superfluid~\cite{Sarma:1963,Alford:1999xc,Liu:2002gi,Shovkovy:2003uu,Alford:2003fq};
phase separation (PS) between the BCS and normal components~\cite{Bedaque:2003hi,Caldas:2004}; a ``magnetized'' paired superfluid ($\rm SF_M$)~\cite{Leo}, and the elusive Larkin, Ovchinnikov, Ferrel and Fulde~(LOFF) phase, in which pairing may occur with a spatially varying superfluid order parameter~\cite{Fulde:1965,Larkin:1965}.

It has been shown that in weak-coupling an asymmetry between the density of the two spin-species results in phase separation both in three~\cite{Bedaque:2003hi,Caldas:2004} and in two-dimensions~\cite{Caldas:2012}: a mixed (heterogeneous) phase, composed by a superfluid paired core surrounded by a shell of expelled normal unpaired fermions. Nevertheless, experiments on trapped population imbalanced Fermi gases are mostly focused on the unitary regime $-1 < 1/k_F a < 1$. Indeed, experiments in the strongly interacting regime have observed phase separation by two independent groups~\cite{Hulet:2006}, and \cite{Shin:2006,Zwierlein:2006}. Then, the theoretical investigation of PS in strong coupling is not only of academic interest. Our main concern here is the exploration of phase separation in a spin-polarized system beyond mean-field. To this aim, we take into account the leading order $k_F a$ corrections to both normal and BCS free-energies and as a consequence, all relevant quantities of interest carry this dependency. We show that PS is stable against the normal phase in the whole BCS range $-\infty < 1/k_F |a| < 0$. We have calculated the magnetization of a partially polarized normal Fermi gas and also in the normal region of the PS state. We also employ a phenomenological approach to describe the observed~\cite{NFL:2010} superfluid-normal transition of a phase-separated Fermi gas at unitarity in terms of the number imbalance $\delta n$.

We also verified the consequences on the Chandrasekhar-Clogston limit after the consideration of the leading order $k_Fa$ corrections to the grand potentials of the normal and BCS phase. We find analytical expressions for the ratio of the critical chemical potential imbalance and the pairing gap, and the critical polarization to leading order in the interaction parameter $k_Fa$, showing a clear improvement of standard mean-field results.

For completeness, we have also investigated the stability of a Bose-Fermi mixture in the far-BEC limit. We find that the molecular BEC is locally stable against an external magnetic field $h$, provided $|h|<\Delta$.

The paper is organized as follows. In Sec.~\ref{MH} we present the model Hamiltonian describing the system of interest, and provide some basic definitions. In this section we investigate the conditions of equilibrium between the normal and superfluid phases, of the phase-separated state, and find the critical chemical potential imbalance and the critical polarization (at which superfluidity is disrupted), both corrected with the first-order $k_F|a|$ correction. In addition, we shall present a zero temperature phase diagram in the $\delta \mu_c/\Delta - k_F |a|$ plane. In this section, we also investigated the stability of a Bose-Fermi mixture in the BEC limit. In Sec.~\ref{comp} we compare our results with previous related work. We conclude in Sec.~\ref{conc}.

\section{Model Hamiltonian and Methods}
\label{MH}

We consider a zero temperature homogeneous (i.e., in the absence of a trapping potential) two-component Fermi system in three-dimensions (3d), consisting of non-relativistic spin-up and spin-down fermions at finite polarization, whose Hamiltonian is given by
\begin{eqnarray} 
H&=&\sum_{k,s=\uparrow,\downarrow}  {\varepsilon}^{s}_k a_{k,s}^\dagger a_{k,s}+ g\sum_{k,p,q} a_{k+q\uparrow}^\dagger a_{p-q\downarrow}^\dagger
a_{k\uparrow} a_{p\downarrow},
\end{eqnarray}
where $g$ is an effective four-fermion contact interaction whose strength at low energy is completely controlled by the two-body scattering length $a$, ${\varepsilon}^{s}_k = \frac{\hbar^2 k^2}{2m} -
\mu_{s}$ is the single-particle dispersion relation for spin species $s$ and momentum $k$, $m$ is the fermion mass, and $a^{\dagger}_{k,s}$, $a_{k,s}$ are the creation and annihilation operators for the spin-$\uparrow$ particles (and the same for the spin-$\downarrow$ particles). To represent an attractive s-wave interaction between the spin-$\uparrow$ and spin-$\downarrow$ atoms we take $g < 0$. From now on we set $\hbar =1$.

For the two component system, the spin-up and spin-down chemical potentials may be written as $\mu_{\uparrow} = \mu + \delta \mu$ and $\mu_{\downarrow} = \mu - \delta \mu$, respectively, where $\delta \mu = \frac{\mu_{\uparrow} - \mu_{\downarrow} }{2}\equiv h$ plays the role of an effective external (Zeeman) magnetic field. The density $n=n_{\uparrow}+n_{\downarrow}$ determines $\mu$ and the polarization density $\delta n=n_{\uparrow}-n_{\downarrow}$ determines $\delta \mu$.  In trapped atom experiments this (imbalanced) regime is reached by calibrating the population of the spin-up and spin-down species.

There is a solid amount of material describing the crossover from the BCS regime of long-range Cooper pairs to the BEC regime of tightly bound molecules, both theoretical and experimental. See, for instance, references~\cite{Review,Zwerger}. We shall focus on the phase separation in the BCS regime and on the Bose-Fermi mixture in the BEC regime.

\subsection{Results for the BCS side}

The ``BCS limit'' on the BCS regime $a<0$, and $\mu>0$, corresponds to $k_F|a| \ll 1$. In previous works it was not considered the inclusion of the leading order $k_F a$ corrections to the thermodynamic potential~\cite{Bedaque:2003hi,Caldas:2004}. Considering these corrections~\cite{Fetter:1971,Sanjay:2005}, the (mean-field) grand-canonical thermodynamic potential or free energy of the normal and BCS states are given, respectively, by

\begin{eqnarray}
\label{OmegaN}
\Omega^{\rm Normal}(\mu_{\downarrow},\mu_{\uparrow})=&-&\frac{k_{F\uparrow}^5}{30 \pi^2 m} - \frac{k_{F\downarrow}^5}{30 \pi^2 m}\\
\nonumber
&-& \frac{a}{9 \pi^3 m}~k_{F\uparrow}^3 k_{F\downarrow}^3,
\end{eqnarray}

\begin{equation}
\Omega^{\rm BCS}(\bar \mu)=-\frac{k_{\rm
  F}^5}{ 15 \pi^2 m} - \frac{mk_{\rm F}}{4\pi^2} \Delta^2 - \frac{a}{9 \pi^3 m} ~k_{\rm F}^6,
\label{BCS}
\end{equation}
where $k_{F\uparrow \downarrow} =  \sqrt{2m\mu_{\uparrow \downarrow}}$, and the zero temperature gap $\Delta$ in the weak-coupling limit, $\Delta / \bar \mu \ll1$, is given by

\begin{equation}
\label{Delta_0}
\Delta(k_{\rm F}a) \approx 4 \bar\mu e^{-2 -\frac{\pi}{2 k_F |a|}},
\end{equation}
with $\bar \mu \equiv  \mu_{\uparrow} + \mu_{\downarrow}$, and $k_F \equiv \sqrt{m\bar\mu} = \sqrt{2m\mu}$. Notice in Eq.~(\ref{OmegaN}) the presence of a term proportional to the s-wave scattering length $a$. The interaction-dependent (Hartree) term is the first beyond mean-field contribution to the mean-field equation of state of the normal phase. As remarked in Ref.~\cite{Pilati:2008}, in order to describe reliably the phase separated state it is crucial to take into account the interaction effects in the normal phase~\cite{Chevy:2006,Aurel:2007}. The normal energy contribution to the BCS phase in Eq.~(\ref{BCS}) also received a first beyond mean-field correction proportional to $k_Fa$.

The number densities in the normal and BCS phase are given, respectively, by $n_{\uparrow \downarrow}= - \frac{\partial \Omega^{\rm Normal}}{\partial \mu_{\uparrow \downarrow}}$ and $n = - \frac{\partial \Omega^{\rm BCS}}{\partial \mu}$, yielding

\begin{equation}
n_\uparrow= \frac{k_{F\uparrow}^3}{6 \pi^2} + \frac{a}{3 \pi^3} k_{F\downarrow}^3 k_{F\uparrow},
\label{numberup}
\end{equation}

\begin{equation}
n_\downarrow= \frac{k_{F\downarrow}^3}{6 \pi^2} + \frac{a}{3 \pi^3} k_{F\uparrow}^3 k_{F\downarrow},
\label{numberdown}
\end{equation}
and

\begin{equation}
\label{numberBCS}
n = \frac{k_{F}^3}{6 \pi^2} + \frac{a k_{F}^4}{3 \pi^3} + \frac{m^2 \Delta^2}{8 \pi^2 k_{F}} \left( 5 + \frac{\pi}{k_{F} |a|} \right).
\end{equation}
From equations~(\ref{numberup}) to~(\ref{numberBCS}) it is clear that the chemical potentials in the normal and BCS phases also receive corrections at order $k_Fa$.

The magnetizations (in the individual and independent phases) are given by $\delta n_{\rm BCS} = - \frac{\partial \Omega^{\rm BCS}}{\partial \delta \mu}$, and $\delta n= - \frac{\partial \Omega^{\rm Normal}}{\partial \delta \mu}$, which give

\begin{equation}
\label{deltanumberBCS}
\delta n_{\rm BCS} = 0,
\end{equation}
as expected, and

\begin{eqnarray}
\delta n= \frac{1}{6 \pi^2} \left[ k_{F\uparrow}^3 \left(1+ \frac{2 k_{F\downarrow} |a|}{ \pi}   \right)  - k_{F\downarrow}^3 \left(1+ \frac{2 k_{F\uparrow} |a|}{ \pi}   \right) \right].
\label{deltanumberN}
\end{eqnarray}
From the above equation, we can obtain a simple expression for the magnetization of the normal phase for the case of small $\delta \mu / \mu$. Expanding Eq.~(\ref{deltanumberN}) in powers of $\delta \mu / \mu$, we find

\begin{eqnarray}
\delta n= \frac{k_F^{3}}{6 \pi^2} \frac{\delta \mu}{\mu} \left[3+ \frac{4 k_F |a|}{ \pi}   \right].
\label{deltanumberN2}
\end{eqnarray}
The facts that the BCS phase is unpolarized and the normal phase is always (partially) polarized for any finite chemical potential asymmetry $\delta \mu$ are well known. What is new here is the $k_F |a|$ correction to the magnetization of the normal phase.

The spin susceptibility $\chi_{N}$, is defined as

\begin{eqnarray}
\chi_{N} = \frac{\partial \delta n}{\partial \delta \mu}= \frac{k_F^{3}}{3 \pi^2} \frac{1}{\mu} \left[1+ \frac{2 k_F |a|}{ \pi}   \right].
\label{magsuscep}
\end{eqnarray}
Thus, we can write the magnetization as 

\begin{eqnarray}
\delta n= \chi_{N}^0 \delta \mu \left[1+ \frac{2 k_F |a|}{ \pi}   \right],
\label{deltanumberN3}
\end{eqnarray}
where $\chi_{N}^0 \equiv \frac{k_F^{3}}{3 \pi^2} \frac{1}{\mu}$ is the standard (without the leading-order $k_F |a|$ correction) susceptibility of the normal phase.

We want to investigate now the cases of fixed particle numbers of the different species since this is the pertinent situation to cold atoms experiments. The (Helmholtz) energy of the normal and superfluid phases are expressed as

\begin{eqnarray}
E^{\rm Normal} &=& \Omega^{\rm Normal} + \mu_\uparrow n_\uparrow + \mu_\downarrow n_\downarrow \equiv E^{\rm N}\\
\nonumber
&=& \frac{k_{F\uparrow}^5}{20 \pi^2 m} + \frac{k_{F\downarrow}^5}{20 \pi^2 m} + \frac{a}{3 \pi^3 m} ~k_{F\uparrow}^3 k_{F\downarrow}^3,
\label{Energy-N}
\end{eqnarray}

\begin{eqnarray}
E^{\rm BCS} &=& \Omega^{\rm BCS} + \mu_\uparrow n_\uparrow + \mu_\downarrow n_\downarrow = \Omega^{\rm BCS} + \mu  n \\
\nonumber
&=& \frac{k_{\rm F}^{5}}{ 10 \pi^2 m} + \frac{2 a k_{\rm F}^6}{9 \pi^3 m}  + \frac{m \Delta^2 k_{F}}{8 \pi^2 } \left( 3 + \frac{\pi}{k_{F} |a|} \right),
\label{Energy-BCS}
\end{eqnarray}
where we have used that in the BCS phase $ n_\uparrow = n_\downarrow = n$ and $ \mu_\uparrow + \mu_\downarrow = \mu$.

Writing the chemical potentials as a function of the respective number densities, and inserting in the above equations, we obtain the normal and BCS energies as a function of the densities in both phases as

\begin{eqnarray}
E^{\rm N} (n_{\uparrow},n_{\downarrow}) &=& \frac{(6 \pi^2 n_{\uparrow})^\frac{5}{3}}{20 \pi^2 m} + \frac{(6 \pi^2 n_{\downarrow})^\frac{5}{3}}{20 \pi^2 m} + \frac{12 \pi a n_{\uparrow} n_{\downarrow}}{ m},
\label{Energy-N}
\end{eqnarray}

\begin{eqnarray}
\label{Energy-BCSn}
E^{\rm BCS} (n) &=& \frac{(6 \pi^2 n)^\frac{5}{3}}{10 \pi^2 m} \left( 1 + \frac{10 |a| (6 \pi^2 n)^\frac{1}{3}}{3 \pi} \right) \\
\nonumber
&-& \frac{2 |a| (6 \pi^2 n)^2}{9 \pi^3 m} \\
\nonumber
&-& \frac{ m \Delta^2 (6 \pi^2 n)^\frac{1}{3}}{4 \pi^2} \left( 1 - \frac{14 |a| (6 \pi^2 n)^\frac{1}{3}}{3 \pi} \right).
\end{eqnarray}

We notice that if we write the concentration of the minority spin-$\downarrow$ atoms as the ratio of the densities $y= n_\downarrow/n_\uparrow$, the expression of the partially polarized normal gas in Eq.~(\ref{Energy-N}) can be expressed as

\begin{eqnarray}
E^{\rm N}(y) &=& E_{\uparrow} \left[ 1 + A y + y^{5/3} \right],
\label{Energy-up}
\end{eqnarray}
where $E_{\uparrow} \equiv \frac{(6 \pi^2 n_{\uparrow})^\frac{5}{3}}{20 \pi^2 m} $ is the ideal gas Fermi energy, and $A\equiv \frac{20 k_{F \uparrow}a}{3 \pi}$, in which the term $A y$ represents the ``binding'' energy of the $\downarrow$ atoms to the Fermi gas of $\uparrow$ atoms~\cite{Lobo}. For $y=0$ the partially polarized normal gas reduces to the fully polarized normal one, $E^{\rm N}(y=0) = E_{\uparrow}$. The other extreme is reached when $y=1$, that corresponds to an unpolarized normal phase, which is unstable against the superfluid phase ($a<0$), giving $E^{\rm N}(y=1) = E_{\uparrow} \left[ 2 + A \right]$.

\subsubsection{Phase separation}
A remarkable feature of Ref.~\cite{Zwierlein:2006} is the observation of phase separation between the normal and superfluid phases in experiments with imbalanced trapped fermionic atoms. We would like to emphasize that as we mentioned before, in this work we address only the situation of a homogeneous configuration i.e., an infinite system without an external trapping potential. 

In the phase separation state, $n_\uparrow$ and $n_\downarrow$ particles are accommodated in a volume $V$ of a ``box'' in such a way that in a fraction $x$ of this volume the particles are ``free'' having densities $\tilde n_{\uparrow}$ and $\tilde n_{\downarrow}$, and in the rest of the volume there is pairing formation between the spin-$\uparrow$ and spin-$\downarrow$ species with number densities $n_\uparrow^{\rm BCS}=n_\downarrow^{\rm BCS}=n$~\cite{Bedaque:2003hi,Caldas:2004}. Then, the number densities in each component of the mixed or heterogeneous phase read

\begin{eqnarray}
\label{eq0}
n_\uparrow =   x\tilde n_\uparrow + (1-x)n,\\
\nonumber
n_\downarrow = x\tilde n_\downarrow + (1-x)n,
\end{eqnarray}
with $0 \leq x \leq 1$. The preferable phase separated state for given $n_\uparrow$ and $n_\downarrow$ particle densities is the one which has the lowest energy

\begin{equation}
\label{eq13}
{E}^{\rm PS}(n_\uparrow,n_\downarrow)=Min_{x, n}\left\{ x { E}^{\rm N}(\tilde n_\uparrow,\tilde n_\downarrow)+ (1-x){ E}^{\rm BCS}( n) \right\},
\end{equation}
where ${ E}^{\rm N}$, the energy of the normal (unpaired) particles, is given by

\begin{eqnarray}
\label{eq1}
{ E}^{\rm N}(\tilde n_a,\tilde n_b)&=&\frac{(6\pi^2)^{\frac{5}{3}}}{20\pi^2 m}  \left(\frac{n_\uparrow-(1-x)n}{x}\right)^{\frac{5}{3}} \\
\nonumber
&+& \frac{(6\pi^2)^{\frac{5}{3}}}{20\pi^2 m} \left(\frac{n_\downarrow-(1-x)n}{x}\right)^{\frac{5}{3}} \\
\nonumber
&+& \frac{12 \pi a }{ m} \left(\frac{n_\uparrow-(1-x)n}{x}\right) \left(\frac{n_\downarrow-(1-x)n}{x}\right),
\end{eqnarray}
and ${ E}^{\rm BCS}(n)$ is given by Eq.~(\ref{Energy-BCSn}). At $x=0$ in Eqs.~(\ref{eq0}) and~(\ref{eq13}), respectively, the whole system is a conventional BCS superfluid, with $n_\uparrow = n_\downarrow = n$, and at $x=1$ the entire system is in the normal phase with $n_\uparrow =   \tilde n_\uparrow$ and $n_\downarrow = \tilde n_\downarrow$. 

In Eq.~(\ref{eq1}) we neglected the surface energy~\cite{Mueller:2006,Caldas:2007} at the interface between the BCS and normal phases, since this term is negligible in the thermodynamic limit considered here. The surface energy contribution may be important in describing experiments on highly elongated traps, which provide some evidence for the breakdown of the local density approximation~\cite{Mueller:2006,Marchetti:2007}.

In order to obtain an analytic expression for ${ E}^{\rm PS}(n_\uparrow,n_\downarrow)$, we consider that $n_\uparrow = n_\downarrow + \delta n$, where the ``magnetization'' $\delta n$ is assumed to be small, i.e., $\delta n \ll n_\downarrow$. Besides, we set the density of the superfluid component of the PS as $n = n_\downarrow$. This immediately gives for the densities which enter Eq.~(\ref{eq1}), $\tilde n_\downarrow = n_\downarrow$, and $\tilde n_\uparrow = n_\downarrow + \frac{\delta n}{x}$. Then, after expanding Eq.~(\ref{eq1}) up to second order in $\delta n/n_\downarrow$, the expression for ${ E}^{\rm N}(\tilde n_a,\tilde n_b)$ turns out to be

\begin{eqnarray}
\label{NormalEnergy}
\nonumber
{ E}^{\rm N}(\tilde n_a,\tilde n_b)&=&\frac{(6\pi^2 n_\downarrow)^{\frac{5}{3}}}{10\pi^2 m}  \left[1+ \frac{5}{6} \frac{\delta n}{n_\downarrow x} + \frac{5}{18} \left(\frac{\delta n}{n_\downarrow x}\right)^2 \right] \\
&+& \frac{12 \pi a n_\downarrow^2}{m} \left[1+ \frac{\delta n}{n_\downarrow x} \right].
\end{eqnarray}

\begin{figure}[ht]
\includegraphics[height=6cm,width=8cm]{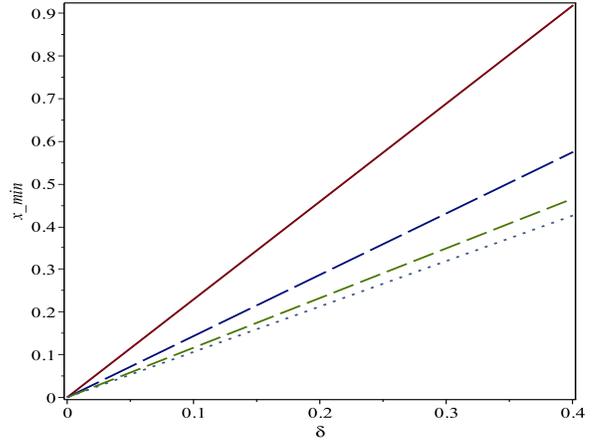}
\caption{(Color online). The volume fraction $x_{min}$ of the normal phase in the PS as a function of $\delta = \delta n /n_{\downarrow}$ for various values of $k_F |a|/\pi$. The top curve is the weak-coupling result. The other ones, from top to bottom, are for $(6 \pi^2 n_\downarrow)^{\frac{1}{3}}|a|/\pi = 0.4, 0.6$, and $0.8$.}
\label{xmin}
\end{figure}

After inserting Eqs.~(\ref{NormalEnergy}) and~(\ref{Energy-BCSn}) in Eq.~(\ref{eq13}), the minimization of ${ E}^{\rm PS}(n_\uparrow,n_\downarrow)$ with respect to $x$ gives

\begin{eqnarray}
\label{MinimumX}
x_{min} = \left[\frac{\frac{(6 \pi^2 n_\downarrow)^{\frac{5}{3}}}{36 \pi^2 m} }{F(k_{F\downarrow}|a|)+G(\Delta(k_{F\downarrow}|a|))}\right]^{\frac{1}{2}} \frac{\delta n}{n_\downarrow},
\end{eqnarray}
where $F(k_{F\downarrow}|a|) \equiv -\frac{ |a|}{9 \pi^3 m}(6 \pi^2 n_\downarrow)^{2}-\frac{12 \pi |a| n_\downarrow^2}{m}=-\frac{4|a|(6 \pi^2 n_\downarrow)^{2}}{9 \pi^3m}$, and $G(\Delta(k_{F\downarrow}|a|)) \equiv \frac{m}{4 \pi^2} \Delta(k_{F\downarrow}|a|)^2 (6 \pi^2 n_\downarrow)^{\frac{1}{3}} \left(1- \frac{14 |a| (6 \pi^2 n_\downarrow)^{\frac{1}{3}} }{3 \pi} \right)$. In Fig.~\ref{xmin} we show $x_{min}$ versus $\delta \equiv \delta n /n_{\downarrow}$ for several values of $(6 \pi^2 n_\downarrow)^{\frac{1}{3}}|a|/\pi$. The upper (solid) curve is without the $k_F|a|$ corrections. The next three curves (long dashed, dashed and dotted), are for $(6 \pi^2 n_\downarrow)^{\frac{1}{3}}|a|/\pi = 0.4, 0.6$, and $0.8$, respectively. These results show that for a same given imbalance $\delta n /n_{\downarrow}$ the greater the value of $(6 \pi^2 n_\downarrow)^{\frac{1}{3}}|a|/\pi$ the smaller the volume fraction $x_{min}$ occupied by the normal phase. 

Finally, we can write ${E}^{\rm PS}$ in terms of $x_{min}$ as

\begin{eqnarray}
\label{PS}
\nonumber
&&{E}_{x_{min}}^{\rm PS}(n_\uparrow,n_\downarrow)=\frac{(6\pi^2 n_\downarrow)^{\frac{5}{3}}}{10\pi^2 m}  \left[1+ \frac{5}{6} \frac{\delta n}{n_\downarrow } + \frac{10}{36} \left(\frac{\delta n}{n_\downarrow}\right)^2 \right] \\
&+& \frac{12 \pi a n_\downarrow^2}{m} \left[1+ \frac{\delta n}{n_\downarrow } \right]\\
\nonumber
&-&(1-x_{min})^2 \left[ F(k_{F\downarrow}|a|)+G(\Delta(k_{F\downarrow}|a|)) \right].
\end{eqnarray}
We can also find the energy difference between the PS and the normal phase, $\Delta E \equiv {E}_{x_{min}}^{\rm PS} - { E}^{\rm N}$, which is given by

\begin{eqnarray}
\label{EDif}
\Delta E&=& -\frac{1}{x_{min}^2}(1-x_{min})^2 \frac{(6 \pi^2 n_\downarrow)^{\frac{5}{3}}}{36 \pi^2 m} \left(\frac{\delta n}{n_\downarrow}\right)^2\\
\nonumber
&=&-(1-x_{min})^2 \left[ F(k_{F\downarrow}|a|)+G(\Delta(k_{F\downarrow}|a|)) \right].
\end{eqnarray}
In the equation above we have made use of Eq.~(\ref{MinimumX}). This last form tells us that we can analyze this result in two different ways. The first one is that of fixed $n_\downarrow$ and $|a|$ so that we can define a non-dimensional energy difference as

\begin{eqnarray}
\label{EDif2}
\tilde \Delta E &\equiv& \frac{\Delta E}{ F(k_{F\downarrow}|a|)+G(\Delta(k_{F\downarrow}|a|)) }\\
\nonumber
&=&-(1-x_{min})^2 <0.
\end{eqnarray}
which is a function only of $\delta n/n_\downarrow$ (see Eq.~(\ref{MinimumX})), the same trend found at weak coupling~\cite{Bedaque:2003hi,Caldas:2004}. The second one is that of fixed $\delta n/n_\downarrow$, which means that $ \Delta E$ in Eq.~(\ref{EDif}) do depend on $k_{F\downarrow}|a|$ (i.e., $(6 \pi^2 n_\downarrow)^{\frac{1}{3}}|a|$), differently from the weak coupling results~\cite{Bedaque:2003hi,Caldas:2004}. In both cases $\Delta E$ will be $<0$ while $G(\Delta(k_{F\downarrow}|a|))+F(k_{F\downarrow}|a|)>0$, or

\begin{eqnarray}
\label{Bigger}
\frac{ m}{4 \pi^2} \Delta(k_{F\downarrow}|a|)^2 (6 \pi^2 n_\downarrow)^{\frac{1}{3}} \left(1- \frac{14 |a| (6 \pi^2 n_\downarrow)^{\frac{1}{3}} }{3 \pi} \right)\\
\nonumber
 -  \frac{(6 \pi^2 n_\downarrow)^{\frac{5}{3}}}{ \pi^2m} \frac{4|a|(6 \pi^2 n_\downarrow)^{\frac{1}{3}}}{9 \pi} >0.
\end{eqnarray}
We have verified numerically that this condition is satisfied for all values of $k_{F\downarrow}a$ in the BCS regime. This means that to first order in $k_F a$, and small $\delta n$ we are considering here, PS is stable and robust in the wide range $-\infty < 1/k_{F\downarrow}a < 0$.

\subsubsection{Chandrasekhar-Clogston limit}

The presence of a spin imbalance between the spin-up and spin-down species necessarily brings about the presence of two Fermi surfaces, which makes pairing difficult. When the imbalance between the two Fermi surfaces is large enough, superfluidity is broken apart and the system undergoes a quantum first-order phase transition toward the normal state. The existence of such a transition at a critical value of the polarization was first proposed by Clogston~\cite{Clogston} and Chandrasekhar~\cite{Chandrasekhar}, in the context of conventional superconductivity. This is known in the literature as the Chandrasekhar-Clogston (CC) limit of superfluidity.

Let us now verify how the CC limit is modified with the consideration of the leading order $k_Fa$ corrections to the thermodynamic potentials in Eqs.~(\ref{OmegaN}) and (\ref{BCS}).

To find an analytical expression for the critical chemical potential imbalance $\delta \mu(k_Fa) = (\tilde\mu_{\uparrow}(k_Fa) - \tilde\mu_{\downarrow}(k_Fa))/2$ at which superfluidity is destroyed is a rather involved problem. The interaction dependent chemical potentials are given as $\mu_{\uparrow\downarrow}(k_Fa) \equiv \tilde \mu_{\uparrow\downarrow} = \mu_{\uparrow\downarrow} + \frac{2a}{3\pi m} \mu_{\downarrow\uparrow}^{3/2}$~\cite{Sanjay:2005}. As a first approximation, to take into account both $h$ and $k_Fa$, we assume $\tilde \mu_{\uparrow\downarrow} = \mu_{\uparrow\downarrow}(h) - \frac{2a}{3\pi m} \mu_{\downarrow\uparrow}^{3/2}(h)$, such that  

\begin{equation}
\label{muupnew}
\tilde \mu_{\uparrow\downarrow} = \mu[1 \pm h/\mu(1 + 2 k_F a/\pi)-4 k_F a/3\pi].
\end{equation}

In order to obtain closed functions for the the N and BCS pressures as a function of these ``renormalized'' chemical potentials, we expand $\tilde \mu_{\uparrow\downarrow}$ in $\Omega^{\rm N}$ and $\Omega^{\rm BCS}$ up to order $k_Fa$ and $(h/\mu)^2$.

\begin{figure}[ht]
\includegraphics[height=7cm,width=8cm]{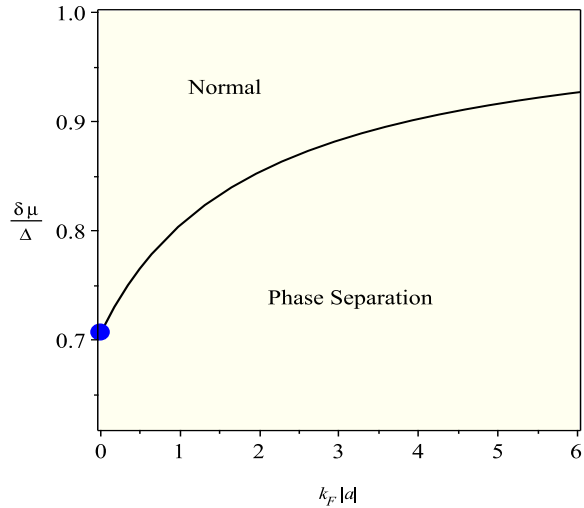}
\caption{(Color online). The zero temperature phase diagram of an imbalanced gas of fermionic atoms in $3d$. The vertical axis (with $k_F |a|=0$) shows that the mean-field ratio of the critical chemical potential imbalance to the pairing gap happens at the CC value (solid blue circle) $\delta \mu_c/\Delta=1/\sqrt{2}$. Taking into account the first-order $k_F |a|$ correction, the critical ratio $\delta \mu_c/\Delta$ increases with $k_F |a|$.}
\label{deltamu}
\end{figure}

Thus, the Gibbs conditions of equilibrium between the normal and the superfluid phase, $-\Omega^{\rm N}=-\Omega^{\rm BCS}$, and $\mu^{BCS}\equiv \tilde \mu=(\tilde \mu_{\uparrow} + \tilde \mu_{\downarrow})/2=\mu(1-4k_Fa/3\pi)$~\cite{Bedaque:2003hi,Caldas:2004}, give

\begin{eqnarray}
\nonumber
& P_0&\left\{2- \frac{20 k_Fa}{3\pi} + \frac{15}{4} \left(\frac{h}{\mu} \right)^2\left(1+ \frac{4 k_Fa}{\pi} \right) \right. \\ 
\nonumber
&+& \left. \frac{10 k_Fa}{3\pi} \left[1-\frac{3}{2} \left(\frac{h}{\mu} \right)^2 \right]  \right\}\\
\nonumber
&=& P_0 \left(1- \frac{10 k_Fa}{3\pi} \right) \left[2 + \frac{10 k_Fa}{3\pi} + \frac{30}{16}\frac{\Delta^2}{\tilde \mu^2} \right],
\label{CC-1}
\end{eqnarray}
where $P_0 \equiv (2m\mu)^{5/2}/30\pi^2m$, whose solution (neglecting the term of order $(k_F a)^2$) is

\begin{eqnarray}
\label{CC}
\frac{\delta \mu_c}{\Delta}&=&\frac{1}{\sqrt{2}}\frac{\sqrt{1+\frac{8 k_F|a|}{3\pi}}}{\sqrt{1+\frac{4 k_F|a|}{3\pi}}},
\end{eqnarray}
where again, $\delta \mu = h= (\mu_{\uparrow} - \mu_{\downarrow})/2$. Eq.~(\ref{CC}) allows us to construct the phase diagram of the $3d$ imbalanced Fermi gas. We show in Fig.~\ref{deltamu} the behavior of the ratio $\delta \mu_c/\Delta$, corrected with the first-order $k_F |a|$ correction, as a function of $k_F |a|$. The resulting curve agrees with previous investigation~\cite{Sanjay:2005}, which showed for the first time an increase in $\delta \mu_c/\Delta$ with increasing $k_F |a|$.

\begin{figure}[ht]
\includegraphics[height=6cm,width=8cm]{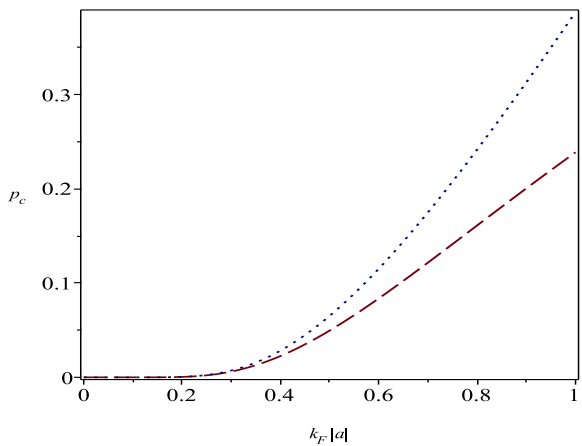}
\caption{(Color online). The critical polarization $p_c$ for the imbalanced Fermi gas as a function of the interaction parameter $k_F |a|$. The top curve shows the mean-field corrected with the first-order $k_F |a|$ correction, while the lower curve is the mean-field result.}
\label{polarizations}
\end{figure}

From Eq.~(\ref{deltanumberN2}) it is very easy to see that the number density difference $\delta n$ can be expressed as,

\begin{equation}
\label{deltan}
\delta n=3 n \frac{\delta \mu}{\mu} \left(1+\frac{4k_F|a|}{3\pi} \right).
\end{equation}
This expression can be written in terms of the polarization $p$, defined as

\begin{equation}
\label{polarization}
p = \frac{ n_\uparrow - n_\downarrow}{n_\uparrow + n_\downarrow} = \frac{\delta n}{n}.
\end{equation}
Combining equations~(\ref{CC}),~(\ref{deltan}) and~(\ref{polarization}) yields the critical polarization $p_c$, the value of the polarization at which the transition normal-phase separation occurs

\begin{eqnarray}
\label{critical}
p_c &=& \frac{3}{ \sqrt{2}} \left(\frac{2}{e} \right)^2 e^{ -\frac{\pi}{2 k_F |a|}}\\
\nonumber
&\times& \sqrt{1+\frac{8 k_F|a|}{3\pi}} \sqrt{1+\frac{4 k_F|a|}{3\pi}}.
\end{eqnarray}
In an earlier result obtained by Pilati and Giorgini~\cite{Pilati:2008}, subjected to the same equilibrium conditions, it was found $p_c = 3/ \sqrt{8} \left(2/e \right)^{7/3} e^{ -\frac{\pi}{2 k_F |a|}}$, to leading order in $p_c$. While $p_c$ in Eq.~(\ref{critical}) is obviously not expected to be valid at unitarity, it is clearly an improvement over the mean-field result $p_c^0$, where $p_c^0 = \frac{3}{ \sqrt{2}} \left(\frac{2}{e} \right)^2 e^{ -\frac{\pi}{2 k_F |a|}} $ is obtained from the standard CC limit $\delta \mu=\frac{\Delta}{\sqrt{2}}$. Notice that a purely mean-field result at unitarity predicts $p_c = 0.93$~\cite{Leo2}, whereas in Ref.~\cite{Lobo} they found $p_c = 0.77$, and in Ref.~\cite{Pilati:2008} it is found $p_c = 0.39$, both results obtained by means of quantum Monte Carlo simulations. Experiments found $p_c \approx 0.75$ at the Feshbach resonance, and $p_c \approx 0.52$ for $k_Fa \approx -3.7$ (at the BCS side)~\cite{Zwierlein:2006}. In Fig.~\ref{polarizations} we show the behavior of the critical polarization $p_c$ for the imbalanced Fermi gas as a function of the interaction parameter $k_F |a|$. The lower (dashed) curve is the mean-field result, while the top (doted) curve shows the mean-field corrected to leading order in $k_F |a|$.

\subsubsection{Magnetizations}

The magnetization in the BCS and normal phases of the PS can be obtained, respectively, as

\begin{equation}
\label{deltanumberBCS2}
M^{\rm BCS} =0,
\end{equation}
by the construction of the PS state (see Eq.~\ref{eq0}). Notice that as pointed out in Ref.~\cite{NFL:2010}, {\it the superfluid phase is not polarized, whatever the chemical potential imbalance}, while in normal phase the magnetization is given by

\begin{eqnarray}
M^N \propto \tilde n_{\uparrow} - \tilde n_{\downarrow} = C (\tilde n_{\uparrow} - \tilde n_{\downarrow}) = C\frac{\delta n}{x},
\label{deltanumberN3}
\end{eqnarray}
where $C$ is proportional to the normal susceptibility. At the minimum $x=x_{min}$, so that

\begin{eqnarray}
M^N=C \frac{\delta n}{x_{min}}= Cn_{\downarrow} \left[\frac{F(k_{F\downarrow}|a|)+G(\Delta(k_{F\downarrow}|a|))}{\frac{(6 \pi^2 n_\downarrow)^{\frac{5}{3}}}{36 \pi^2 m} }\right]^{\frac{1}{2}}.
\label{deltanumberN4}
\end{eqnarray}

It is worth to notice that observations of the polarization $p$ in the superfluid phase of the unitary Fermi gas show that it remains equal to $0$ and then jump to $p \approx 0.4$ at the superfluid/normal transition (for $\mu_1-\mu_2 \approx 0.4\times2E_F$)~\cite{NFL:2010}. In order to describe this observed behavior of the polarization of a unitary Fermi gas, and since the system is phase separated~\cite{Frederic}, we adopt a phenomenological approach by imposing an ansatz for the number densities in the PS,

\begin{eqnarray}
\label{ansatz}
n_\uparrow =   x\tilde n_\uparrow + (1-x)n^{\rm BCS}_\uparrow,\\
\nonumber
n_\downarrow = x\tilde n_\downarrow + (1-x)n^{\rm BCS}_\downarrow,
\end{eqnarray}
where $n^{\rm BCS}_\uparrow \equiv n+ \theta(\delta n - \delta n_c) \delta n$ and $n^{\rm BCS}_\downarrow \equiv n$. $\theta(x)$ is the Heaviside step function and $\delta n_c$ is a critical value for the number density asymmetry, introduced to represent the jump in $p$ from $0$ to $ \approx 0.4$ at the superfluid/normal transition in~\cite{NFL:2010}. Setting $n=n_\downarrow$ as before, we find from Eq.~(\ref{ansatz}) $\tilde n_\downarrow = n_\downarrow $ and $\tilde n_\uparrow =   n_\downarrow + \frac{\delta n}{x}[1+\theta(\delta n - \delta n_c) (x-1)]$. Thus, the magnetizations in the BCS and normal phase, respectively, now read

\begin{equation}
\label{deltanumberBCS3}
M^{\rm BCS} = C(n^{\rm BCS}_\uparrow - n^{\rm BCS}_\downarrow)=C\theta(\delta n - \delta n_c) \delta n,
\end{equation}
and

\begin{eqnarray}
M^{\rm N}= C( \tilde n_{\uparrow} - \tilde n_{\downarrow} )= C \left[\frac{\delta n}{x} + \theta(\delta n - \delta n_c) \delta n (1-\frac{1}{x})\right],
\label{deltanumberN5}
\end{eqnarray}
which increases linearly with $\delta n$ for $\delta n \geq \delta n_c$, and the same for $M^{\rm BCS}$ i.e., $M^{\rm BCS}=M^{\rm N} \equiv M=C \delta n$ signalizing the ``melting'' of the PS, while for $\delta n < \delta n_c$ the results are given by Eqs.~(\ref{deltanumberBCS2}) and~(\ref{deltanumberN4}). Notice from Eq.~(\ref{deltanumberN4}) that $M^{\rm N}(\delta n < \delta n_c)$ within the PS is a constant for a given (fixed) $n_{\downarrow}$ and $k_F |a|$, and is $\propto 1/\alpha$, where $\alpha$ is the slope of the respective curve $x_{min}$ versus $\delta n/n_{\downarrow}$ in Fig.~\ref{xmin}. Defining now the density in the BCS phase which enter Eq.~(\ref{Energy-BCSn}) as $n^{\rm BCS} \equiv \frac{n^{\rm BCS}_\uparrow + n^{\rm BCS}_\downarrow }{2}$, we find $n^{\rm BCS} = n +\frac{\theta(\delta n - \delta n_c) \delta n}{2}$, and following the same steps as before, we find

\begin{eqnarray}
\label{MinimumX2}
x_{min} = \left[\frac{\frac{(1-\theta(\delta n - \delta n_c)) (6 \pi^2 n_\downarrow)^{\frac{5}{3}}}{36 \pi^2 m} }{\tilde F(k_{F\downarrow}|a|)+\tilde G(\Delta(k_{F\downarrow}|a|))}\right]^{\frac{1}{2}} \frac{\delta n}{n_\downarrow},
\end{eqnarray}
where we have defined $\tilde F(k_{F\downarrow}|a|) =\frac{|a|(6 \pi^2 n_\downarrow)^{2}}{6 \pi^3m}\left(1+\frac{4}{3} f(\theta) \right)$, and $\tilde G(\Delta(k_{F\downarrow}|a|)) = \frac{m}{4 \pi^2} \Delta(k_{F\downarrow}|a|)^2 (6 \pi^2 n_\downarrow)^{\frac{1}{3}} \left(1- \frac{14 |a| (6 \pi^2 n_\downarrow)^{\frac{1}{3}} }{3 \pi} \right) \left(1+\frac{f(\theta)}{3} \right)$, with $f(\theta) \equiv \frac{\delta n}{n_{\downarrow}} \theta(\delta n - \delta n_c)$. Notice that at $\delta n_c$ i.e., at the superfluid-normal transition, $x_{min} \to 0$ since there is only one (a normal homogeneous) phase now.

\begin{figure}[ht]
\includegraphics[height=6cm,width=8cm]{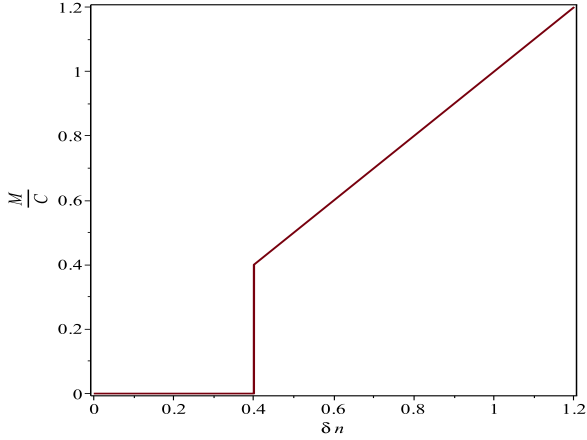}
\caption{The reduced polarization $M^{{\rm BCS}}/C$ for $\delta n < \delta n_c$ and $M/C$ for $\delta n \geq \delta n_c$ for the imbalanced Fermi gas in the unitary regime as a function of the density asymmetry $\delta n$. The curve shows a jump at $\delta n_c =0.4$ signalizing the first-order quantum superfluid/normal phase transition~\cite{He}, as observed in~\cite{NFL:2010}.}
\label{polarization1}
\end{figure}

In Fig.~\ref{polarization1} we show the ``reduced'' polarizations $M^{{\rm BCS}/{\rm N}}/C$ for the imbalanced Fermi gas in the unitary regime as a function of the density asymmetry $\delta n$.

\subsection{Results for the BEC side}

On the BEC side of the resonance, $a>0$, and $\mu<0$, where the ``BEC limit'' is characterized by $k_Fa \ll 1$, the mean-field equation of state describing a mixture of $n_b=n_{\downarrow}$ bosonic dimers and $n_a = n_\uparrow - n_\downarrow $ unpaired fermionic atoms~\cite{Viverit,Viverit2}, and chapter 11 of \cite{Zwerger}, is

\begin{widetext}
\begin{eqnarray}
E^{BF} =  \overbrace{E(\epsilon_b)}^{ binding~energy~of~dimers} +\overbrace{ \frac{g_{bb} n_b^2}{2}}^{dimer-dimer~interaction} +\overbrace{g_{ab} n_a n_b}^{atom-dimer~interaction} +\overbrace{ E_a}^{energy~of~unpaired~atoms},
\label{BF}
\end{eqnarray}
\end{widetext}
where $E(\epsilon_b) = n_b \epsilon_b$, $\epsilon_b= -\frac{1}{m a^2}$ is the binding energy of dimers, $E_a= \frac{(6 \pi^2 n_a)^{\frac{5}{3}}}{2m} = \frac{3}{5} n_a E_{Fa}$, and where $E_{Fa}=\frac{k_{Fa}^2}{2m}=\frac{(6 \pi^2 n_a)^{\frac{2}{3}}}{2m}$ is the Fermi energy of the remaining (unpaired) atoms. Recalling the standard definitions of the interactions in the equation above in terms of the scattering lengths $g_{bb}=\frac{4\pi a_{bb}}{m_b}$, where $m_b=2m$ is the dimer mass and $a_{bb}$ is the dimer-dimer scattering length, which is assumed as positive, $g_{ab}=\frac{4\pi a_{ab}}{m_{ab}}$, where $m_{ab}=2m m_b/(m+m_b)$ and $a_{ab}$ is the atom-dimer scattering length. $a_{bb}$ and $a_{ab}$ are given in terms of the two-body s-wave scattering length $a$ as $a_{bb}=0.6a$~\cite{Petrov,Petrov2}, and $a_{ab}=1.2a$~\cite{Skorniakov}. Thus, Eq.~(\ref{BF}) can be rewritten as

\begin{eqnarray}
E^{BF} &=&  (\epsilon_b+ g_{ab} n_a) n_b + \frac{g_{bb} n_b^2}{2} + E_a.
\label{BosonFermion}
\end{eqnarray}
In order to obtain the grand potential $\Omega = E - \mu_a n_a - \mu_b n_b$, which is useful to investigate the CC limit in the boson-Fermi mixture, we follow an interesting analysis in chap. 11 of~\cite{Zwerger}. From the equation above we find the chemical potentials $\mu_a$ and $\mu_b$,

\begin{eqnarray}
\label{XX}
\mu_b &=& -|\epsilon_b| + g_{ab} n_a + g_{bb} n_b =  \mu_{\uparrow} + \mu_{\downarrow} = 2 \bar \mu,\\
\nonumber
\mu_a &=& g_{ab} n_b + E_{Fa} = \mu_{\uparrow} =  \bar \mu +h,
\end{eqnarray}
where, as usual, $ \bar \mu \equiv (\mu_{\uparrow} + \mu_{\downarrow})/2$, and $h \equiv (\mu_{\uparrow} - \mu_{\downarrow})/2$ is the same as defined before, and represents an effective external Zeeman magnetic field. Then,

\begin{eqnarray}
\Omega &=&  \Omega(n_b) - (|h| + \bar \mu - g_{ab} n_b) n_a  + E_a.
\label{Potnew}
\end{eqnarray}
where $\Omega(n_b)= \epsilon_b n_b + \frac{g_{bb} n_b^2}{2} - 2 \bar \mu n_b$ is the pure (unpolarized) superfluid potential. It is convenient to introduce the gap $\Delta_{gap} = \frac{|\epsilon_b|}{2} + (g_{ab} - \frac{g_{bb}}{2})n_b $~\cite{Alzetto}, which corresponds to one-half of the energy required to break a pair~\cite{Pitaevskii}. 

After the introduction of $\bar \mu$ from Eq.~(\ref{XX})-a in Eq.~(\ref{Potnew}), it is straightforward to verify that

\begin{eqnarray}
\label{Almost}
\Omega &\equiv& \Omega(n_a,n_b) -  \Omega(n_b)\\ 
\nonumber
&=& - (|h| - \Delta_{gap}) n_a + A n_a^{\frac{5}{3}} - \frac{g_{ab}}{2} n_a^2,
\end{eqnarray}
where $A \equiv \frac{(6 \pi^2)^{\frac{5}{3}}}{2m}$. 

It should be noticed that although $\Omega$ in Eq.~(\ref{Almost}) is not the grand potential yet, it will serve our purposes. Since in Eq.~(\ref{Almost}) and in Eq.~(11.22) of \cite{Zwerger} $\Omega = \Omega(n_a,n_b,\bar \mu,h)$ and the true grand potential $\Omega$ has to be a function of $\bar \mu$ and $h$ only. Strictly speaking one should find $n_a$ and $n_b$ as a function of $\bar \mu$ and $h$ from Eqs.~(\ref{XX}) and plug them in Eq.~(\ref{Almost}). The equations to be solved for $n_a$ and $n_b$ from Eqs.~(\ref{XX}) are

\begin{eqnarray}
\label{Equationnb}
n_b = \frac{(|h| + \bar \mu -  E_{Fa})}{g_{ab}},
\end{eqnarray}
and

\begin{eqnarray}
\label{Equationna}
g_{ab}^2 n_a -g{bb}E_{Fa} - (2\bar \mu - \epsilon_b)g{ab} +  (\bar \mu + |h|)g{bb} =0,
\end{eqnarray}
where $E_{Fa}=A_{Fa} n_a^{\frac{2}{3}}$, with $A_{Fa} \equiv \frac{(6 \pi^2)^{\frac{2}{3}}}{2m}$. Solving Eq.~(\ref{Equationna}) for $n_a$ and inserting $n_a$ in Eq.~(\ref{Equationnb}) one finds $n_b$, where now both $n_{a,b}=n_{a,b}(\bar \mu, h)$, and of course, they will also depend on $g_{ab},g_{bb},\epsilon_b$ and $A$ in the following way: writing Eq.~(\ref{Equationna}) as $ n_a -\frac{g{bb}}{g_{ab}^2} E_{Fa} - (2\bar \mu - \epsilon_b)\frac{1}{g{ab}} +  (\bar \mu + h)\frac{g{bb}}{g_{ab}^2} \equiv n_a - B n_a^{2/3} + C=0$, where $B=\frac{g{bb}}{g_{ab}^2}A_{Fa}$, and $C=(\epsilon_b-2\bar \mu)\frac{1}{g{ab}} +  (\bar \mu + h)\frac{g{bb}}{g_{ab}^2}$, we find three solutions, one real

\begin{eqnarray}
\label{solution}
n_{a,1}=B\left(\frac{1}{6} F(B,C)+\frac{2}{3} \frac{B^2}{F(B,C)}+\frac{1}{3}B \right)^2-C,
\end{eqnarray}
where $F(B,C)=\left(-108C+8B^3+12\sqrt{-12B^3 C+81C^2}\right)^{\frac{1}{3}}$, and the other two solutions, although are similar to the one in Eq.~(\ref{solution}), are not useful since they are complex, and $n_{a,2}=n_{a,3}^*$.

\begin{figure}[ht]
\includegraphics[height=6cm,width=8cm]{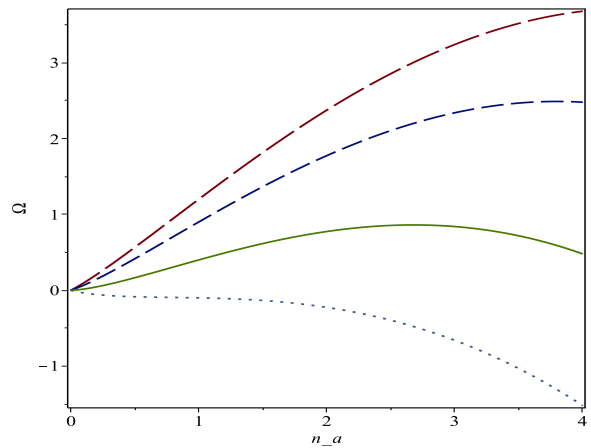}
\caption{(Color online). The grand potential $\Omega=\Omega(n_a,n_b) -  \Omega(n_b)$ of a Bose-Fermi mixture as a function of the unpaired atoms density $n_a$ for several values of $|h| - \Delta_{gap}$. From top to bottom, the first two curves are for $|h| < \Delta_{gap}$, the third curve is for $|h| = \Delta_{gap}$, and the bottom curve is for $|h| > \Delta_{gap}$.}
\label{grandpotential}
\end{figure}

In Fig.~\ref{grandpotential} we show the ``grand potential'' $\Omega=\Omega(n_a,n_b) -  \Omega(n_b)$ in Eq.~(\ref{Almost}) of a Bose-Fermi mixture as a function of the excess (unpaired) atoms with density $n_a$, for various values of $|h| - \Delta_{gap}$. The atom-dimer interaction $g_{ab}$ was taken also as positive ($a_{ab}=1.2a$). From top to bottom, the first two curves (long dashed and dashed) are for $|h| < \Delta_{gap}$, the third curve (solid) is for $|h| = \Delta_{gap}$, and the bottom curve (dotted) is for $|h| > \Delta_{gap}$. A graphical inspection of Fig.~\ref{grandpotential} shows that there is a second-order phase transition from the unpolarized to the polarized superfluid for $|h|>\Delta_{gap}$. This may be due to the mean-field approximation used to describe the Bose-Fermi mixture in the far-BEC limit, and the actual transition may be of first-order~\cite{Zwerger}. Nevertheless, in any of the situations, we confirm previous findings, chapter 11 of~\cite{Zwerger}, that the molecular BEC is locally stable against an external magnetic field on condition that $|h|<\Delta_{gap}$.

\section{Comparison with a Related Work}
\label{comp}

A similar analysis also including interactions in the normal phase to leading order, was carried out by Carlson and Reddy in Ref.~\cite{Sanjay:2005}. In Fig.~1 of Ref.~\cite{Sanjay:2005} the ratio $\delta\mu / \Delta$ of the critical chemical potential difference to the pairing gap is given as a function of the coupling strength $1/k_Fa$. In the deep BCS limit ($k_F |a| \to 0$) this ratio gives $1/\sqrt{2}$, as in Eq.~(\ref{CC}). The results in Fig.~1 of Ref.~\cite{Sanjay:2005} indicate that the ratio $\delta\mu / \Delta$ increases with increasing coupling strength (i.e., $k_F |a| \to \infty$), as we found in Eq.~(\ref{CC}), although our expression is not valid at unitarity.

Regarding the result we obtained, that the unpolarized Bose-Fermi mixture on the BEC side of the Feshbach resonance is stable provided $|h|<\Delta_{gap}$, it is worth to comment on the ``apparent'' disagreement between this result and the one obtained also by Carlson and Reddy in Ref.~\cite{Sanjay:2005}. In Ref.~\cite{Sanjay:2005} they found that the chemical potential difference $h$ is much larger than the gap $\Delta$ deep in the BEC regime. The reason for this supposable discrepancy lies in the different definitions of the gaps in the two cases. In Ref.~\cite{Sanjay:2005} the gap $\Delta$ ``is the corresponding energy in
the superfluid component of the normal-superfluid mixed phase state'', while here $\Delta_{gap}$ is, by definition, one-half of the energy required to break a pair, which also takes into account interactions between unpaired particles and dimers, properly treated at the mean-field level~\cite{Pitaevskii}. 

As mentinoned above, in Ref.~\cite{Sanjay:2005} it was found that in the extreme BEC limit, $h/\Delta \gg 1$, where the gap $\Delta$ in the BEC limit is given by $\Delta_{BEC}=\frac{4\pi a_{ab}}{m_{ab}} n_b = g_{ab} n_b$~\cite{Sanjay:2005}. Thus, we can rewrite  $\Delta_{gap}$ as $\Delta_{gap} = \frac{|\epsilon_b|}{2} + \Delta_{BEC} - \frac{g_{bb}}{2}n_b$. The condition for the stability of the molecular BEC is then $|h|<\frac{|\epsilon_b|}{2} + \Delta_{BEC} - \frac{g_{bb}}{2}n_b$, or $\frac{|h|}{\Delta_{BEC}}<  1+ \frac{|\epsilon_b|}{2\Delta_{BEC}} - \frac{g_{bb}}{2 \Delta_{BEC}}n_b$. After substituting the values of $g_{ab}$, $g_{bb}$, $m_{b}$ and $m_{ab}$ given below Eq.~(\ref{BF}), we find $\frac{|h|}{\Delta_{BEC}}<  \frac{5}{6} + \frac{|\epsilon_b|}{2\Delta_{BEC}}$ or, equivalently, $\frac{h}{\Delta_{BEC}} > -\frac{5}{6} + \frac{1}{2 m a^2} \frac{1}{\Delta_{BEC}} $ which goes to $ \frac{1}{2 m a^2} \frac{1}{\Delta_{BEC}} \gg 1$ in the strongly interacting molecular limit ($a \to 0$), which shows that there is no contradiction.

\section{Conclusions}
\label{conc}

In summary, we have theoretically investigated phase separation in a two-component imbalanced Fermi gas at zero temperature beyond mean-field. Considering a system with $n_\downarrow$ and $n_\uparrow = n_\downarrow + \delta n$ fermionic atoms, with $\delta n \ll n_\downarrow$, and taking into account the leading order $k_F a$ corrections we found that PS is stable against the normal phase in the entire ``BCS'' range $-\infty < 1/k_{F\downarrow}a < 0$. We have calculated the magnetization of a partially polarized normal Fermi gas and in the normal region of the PS state. For completeness, in order to describe qualitatively the superfluid-normal transition of an imbalanced Fermi gas at unitarity, we have calculated the polarization of the BCS and normal phases in the PS. For a certain critical imbalance $\delta n_c$ there is a first-order quantum phase transition from the superfluid to the normal phase with the consequential melting of the PS. 

We have also verified the consequences on the Chandrasekhar-Clogston limit with the consideration of the leading order $k_Fa$ corrections to the thermodynamic potentials of the normal and BCS phase. We find that as a result, the ratio $\delta \mu_c/\Delta$ and the critical polarization $p_c$ received corrections which also depends on the interaction parameter $k_Fa$, showing a clear improvement of previous standard MF results. We have also presented a zero temperature phase diagram for the $3d$ imbalanced Fermi gas in the $\delta \mu_c/\Delta - k_F |a|$ plane, displaying the regions of phase separation and normal phase.

Finally, now on the other side of the resonance, we investigated the stability of a Bose-Fermi mixture in the far-BEC limit, where the interactions can be safely treated by the mean-field approximation. We find that the molecular BEC is locally stable against an external effective magnetic field $h$, provided $|h|<\Delta_{gap}$.
\\

{\it Acknowledgments:} 

I am grateful to F. Chevy and L. He for stimulating conversations. I also wish to thank CNPq and FAPEMIG for partial financial support.

\vspace{-0.2in}

\begin{thebibliography}{10}
\bibitem{Thomas:2004ex}
J.~Kinast, S.~L. Hemmer, G.~M. E., A.~Turlapov and J.~E. Thomas,
\newblock Phys. Rev. Lett. {\bf 92}, 150402 (2004).
\bibitem{Bartenstein:2004}
M.~Bartenstein {\em et~al.},
\newblock Phys. Rev. Lett. {\bf 92}, 120401 (2004).
\bibitem{Chin:2004}
C.~Chin {\em et~al.},
\newblock Science {\bf 305}, 1128 (2004).
\bibitem{Greiner:2004}
M.~Greiner, C.~A. Regal and D.~S. Jin,
\newblock Phys. Rev. Lett. {\bf 94} 070403 (2005).
\bibitem{Review} M. Inguscio, W. Ketterle, and C. Salomon. Ultracold Fermi Gases. {\it Proceedings of the International School of Physics Enrico Fermi}, Course CLXIV, Varenna, (2006).
\bibitem{Theory1} F. Chevy and C. Mora, Rep. Prog. Phys, {\bf 73}, 112401 (2010).
\bibitem{Theory2} K. B. Gubbels and H. T. C. Stoof, Phys. Rept. {\bf 525}, 255 (2013).
\bibitem{Zwerger} W. Zwerger, Ed., {\it The BCS-BEC Crossover and the Unitary Fermi Gas}. vol. 836, Lecture Notes in Physics, Springer, (2012).
\bibitem{Sarma:1963}
G.~Sarma,
\newblock Phys. Chem. Solid {\bf 24}, 1029 (1963).
\bibitem{Alford:1999xc}
M.~G. Alford, J.~Berges and K.~Rajagopal,
\newblock Phys. Rev. Lett. {\bf 84}, 598 (2000).
\bibitem{Liu:2002gi}
W.~V. Liu and F.~Wilczek,
\newblock Phys. Rev. Lett. {\bf 90}, 047002 (2003).
\bibitem{Shovkovy:2003uu}
I.~Shovkovy and M.~Huang,
\newblock Phys. Lett. {\bf B564}, 205 (2003).
\bibitem{Alford:2003fq}
M.~Alford, C.~Kouvaris and K.~Rajagopal,
\newblock Phys. Rev. Lett. {\bf 92}, 222001 (2004).
\bibitem{Bedaque:2003hi}
P.~F. Bedaque, H.~Caldas and G.~Rupak,
\newblock Phys. Rev. Lett. {\bf 91}, 247002 (2003).
\bibitem{Caldas:2004}
H.~Caldas, 
\newblock Phys. Rev. A {\bf 69}, 063602 (2004).
\bibitem{Leo} D. Sheehy and L. Radzihovsky, Phys. Rev. Lett. {\bf 96}, 060401 (2006).
\bibitem{Caldas:2012} H. Caldas, A. L. Mota, R. L. S. Farias, and L. A. Souza, J. Stat. Mech. (2012) P10019.
\bibitem{Fulde:1965}
P.~Fulde and R.~A. Ferrel,
\newblock Phys. Rev. {\bf 135}, A550 (1964).
\bibitem{Larkin:1965}
A.~I. Larkin and Y.~N. Ovchinnikov,
\newblock Sov. Phys. JETP {\bf 20}, 762 (1965).
\bibitem{Hulet:2006} G. B. Partridge, W. Li, R. I. Kamar, Y. A. Liao, R. G. Hulet, Science {\bf 311}, 503 (2006).
\bibitem{Shin:2006} Y. Shin, M. W. Zwierlein, C. H. Schunck, A. Schirotzek, W. Ketterle, Phys. Rev. Lett. {\bf 97}, 030401 (2006). 
\bibitem{Zwierlein:2006} M. W. Zwierlein et al., Science {\bf 311}, 492 (2006); M. W. Zwierlein et al., Nature (London) {\bf 442}, 54 (2006).
\bibitem{NFL:2010} S. Nascimbene, N. Navon, S. Pilati, F. Chevy, S. Giorgini, A. Georges, and C. Salomon, Phys. Rev. Lett. {\bf 106}, 215303 (2011).
\bibitem{He} L. He, and P. Zhuang, Phys. Rev. B {\bf 83}, 174504 (2011).
\bibitem{Pilati:2008} S. Pilati and S. Giorgini, Phys. Rev. Lett. {\bf 100}, 030401 (2008).
\bibitem{Chevy:2006} F. Chevy, Phys. Rev. A {\bf 74}, 063628 (2006).
\bibitem{Aurel:2007} A. Bulgac and M. M. Forbes, Phys. Rev. A {\bf 75}, 031605(R) (2007).
\bibitem{Fetter:1971}
A.~L. Fetter and J.~D. Walecka,
\newblock {\em Quantum Theory of Many-Particle Systems} (McGraw-Hill Inc., 1971).
\bibitem{Lobo} C. Lobo et al., Phys. Rev. Lett. {\bf 97}, 200403 (2006).
\bibitem{Sanjay:2005} J. Carlson and S. Reddy, Phys. Rev. Lett. {\bf 95}, 060401 (2005).
\bibitem{Mueller:2006} T. N. De Silva and E. J. Mueller, Phys. Rev. Lett. {\bf 97}, 070402 (2006).
\bibitem{Caldas:2007} H. Caldas, J. Stat. Mech. {\bf 11}, 11012 (2007).
\bibitem{Marchetti:2007} M. M. Parish, F. M. Marchetti, A. Lamacraft, B. D. Simons, Nature Physics {\bf 3}, 124 (2007).
\bibitem{Clogston} A. M. Clogston, Phys. Rev. Lett. {\bf 9}, 266 (1962).
\bibitem{Chandrasekhar} B. S. Chandrasekhar, Appl. Phys. Lett. {\bf 1}, 7 (1962).
\bibitem{Leo2} D. E. Sheehy and L. Radzihovsky, Phys. Rev. B {\bf 75}, 136501 (2007); Ann. Phys. (N.Y.) {\bf 322}, 1790 (2007).
\bibitem{Frederic} In Ref.~\cite{NFL:2010} they have also observed a unpolarized core surrounded by a polarized outer cloud, as firstly seen in the MIT~\cite{Shin:2006,Zwierlein:2006} and Rice~\cite{Hulet:2006} experiments. F. Chevy, private communication.
\bibitem{Viverit} L. Viverit, C. J. Pethick, and H. Smith, Phys. Rev. A {\bf 61}, 053605 (2000).
\bibitem{Viverit2} L. Viverit and S. Giorgini, Phys. Rev. A {\bf 66}, 063604 (2002).
\bibitem{Petrov} D. S. Petrov, C. Salomon, and G. V. Shlyapnikov, Phys. Rev. Lett. {\bf 93},090404 (2004).
\bibitem{Petrov2} D. S. Petrov, C. Salomon, and G. V. Shlyapnikov, J. Phys. B {\bf 38}, S645 (2005).
\bibitem{Skorniakov} G. V. Skorniakov and K. A. Ter-Martirosian, Sov. Phys. JETP {\bf 4}, 648 (1957).
\bibitem{Alzetto} F. Alzetto and X. Leyronas, Phys. Rev. A {\bf 81}, 043604 (2010).
\bibitem{Pitaevskii}  S. Giorgini, L. P. Pitaevskii, and S. Stringari, Rev. Mod. Phys. {\bf 80}, 1215 (2008).
\end{thebibliography}

\end{document}